\documentstyle[editedvolume,numreferences]{crckapb} 

\input epsfig

\def\tr{{\rm tr}}

\def\half{{\textstyle {1 \over 2}}}

\def\RR{${\rm R}\otimes{\rm R}~$}
\def\NSNS{${\rm NS}\otimes{\rm NS}~$}
\def\RNS{${\rm R}\otimes{\rm NS}~$}
\def\NSR{${\rm NS}\otimes{\rm R}~$}

\def\mh{{\hat \mu}}
\def\nh{{\hat \nu}}

\def\mm{{\hat m}}
\def\nn{{\hat n}}

\def\Rfour{t^8t^8 R^4}
\def\lam16{\lambda^{16}}
\def\thbar {\bar{\theta}}

\def\xxx#1 {{hep-th/#1}}

\def\npb#1(#2)#3 {{\it Nucl. Phys.} {\bf B#1} (#2) #3 } \def\rep#1(#2)#3
{ Phys.
Rept.{\bf #1} (#2) #3 } \def\plb#1(#2)#3{{\it Phys. Lett.} {\bf #1B} (#2) #3}
\def\prl#1(#2)#3{{\it Phys. Rev. Lett.}{\bf #1} (#2) #3}
\def\physrev#1(#2)#3{{\it Phys. Rev. }{\bf D#1} (#2) #3} \def\ap#1(#2)#3{Ann.
Phys. {\bf #1} (#2) #3} \def\rmp#1(#2)#3{{\it Rev. Mod. Phys.}{\bf #1}
(#2) #3}

\begin{opening}
\title{Aspects of D-Instantons}
\subtitle{Talk at Nato ASI, Carg{\`e}se 1997, PUPT-1752 , hep-th/9712156}
 
\author{Michael Gutperle}
\institute{Joseph Henry Laboratories,\\
Princeton University,\\ Princeton, New Jersey 08544,  USA}

\end{opening}
\runningtitle{D-Instantons}

\begin{document}

\begin{abstract}
An overview over effects of D-instantons in ten dimensional IIB superstring theory is given, including the supergravity instanton solution, instanton induced effective interaction vertices, the conjectured $SL(2,Z)$ invariant completion of such terms and the connection of such terms to a one loop calculation in eleven dimensional supergravity.
\end{abstract}

\section{Introduction}
D-instantons \index{D-instanton} are responsible for several important  nonperturbative effects in superstring theory.

 Firstly, they appear as euclidian Dirichlet  p-branes whose $p+1$ dimensional worldvolume wraps a supersymmetric  $p+1$ cycle in the compactification manifold. Such configurations were discussed as euclidian membranes and fivebranes in Calabi-Yau compactifications of M-theory \cite{bbstr}, instanton corrections to the hypermultiplet geometry near the conifold in IIA  on a Calabi-Yau manifold  \cite{ooguri}, D-string instanton corrections to $R^4$ and $F^4$ terms in type I string theory which are important for type I/heterotic duality \cite{kiritsis,bachasvanhove}, threshold corrections to $R^4$ terms  and U-duality in toroidally compactified type II theories \cite{kiritsisb}.

Secondly, even in uncompactified ten dimensional IIB superstring theory they make an appearance as $p=-1$ branes \cite{mbgb}, i.e D-branes with worldvolumes which are points in  spacetime.

Indeed in the context of bosonic string theory such D-instantons were the first 'Dirichlet' objects considered in the past. The introduction of boundaries where the embedding coordinates satisfy Dirichlet boundary conditions in all directions was used to
 implement point-like partonic behavior in dual model amplitudes which is  a desirable feature for a string theory of strong interactions  \cite{MBGQCD}. The fact that momentum is not conserved in the presence of Dirichlet boundaries was used to define off-shell continuations of string scattering amplitudes \cite{offshell}.
 It was also observed that the characteristic  softness \cite{grossmende} of string amplitudes in the high energy fixed angle scattering regime, does not hold in the presence of D-instantons \cite{hard}. 
 Polchinski \cite{pola} showed in the context of bosonic strings that the new singularities which arise because of the  pointlike nature of the open string boundary conditions  cancel in a generalization of the Fischler-Susskind mechanism. In this scheme 
 the combinatorics of string perturbation theory in the presence of D-instantons changes  and  one has to include disconnected worldsheets of different topology whose boundaries are mapped into the same spacetime point.

In the context of IIB superstrings D-instantons have
 several new  properties. A D-instanton is a BPS solution \index{BPS states} which preserves half of the spacetime supersymmetries. In \cite{ggp} the supergravity solution 
corresponding to a D-instanton was found. The broken supersymmetries induce fermionic zero modes which have to be integrated over in order to restore spacetime supersymmetry.
 This induces new effective (t'Hooft) vertices in the low energy effective action. In \cite {greengut}  the instanton induced interactions  were  derived and  the $SL(2,Z)$ duality symmetry  of IIB superstring theory was  used to conjecture 
a modular invariant completion  of the $R^4$ term which contains tree level, one loop and instanton corrections. In \cite{greenhove,ggv,ggk} these expressions were  related to similar terms in M-theory and in particular it was argued that they arise from  a one 
loop calculation in eleven dimensions \index{M-theory}. The material presented in this talk  was obtained in collaboration with M.B. Green,
 H.H. Kwon and P. Vanhove and the interested reader is referred to the original papers for a more detailed discussion.

\section{IIB supergravity}
The massless fields of IIB superstring theory \index{IIB supergravity} are given by the bosonic fields : a complex scalar $\rho$, two second rank  AST potentials $B_a^{(2)}$ $a=1,2$, the zehnbein $e^a_\mu$ and a fourth rank AST potential $C^{(4)}$ with selfdual field strength and the fermionic fields:
 a complex spin $1/2$ dilatino $\lambda $ and a complex spin $3/2$ gravitino $\psi_\mu$ of opposite chirality. As usual in  extended supergravity theories  the scalar fields parametrize a  coset, which for IIB supergravity is  $SL(2,R)/U(1)$. Under $SL(2,R)$ transformations only the second rank
 AST transform as  doublets. The fermions $\lambda$ and $\psi_\mu$ transform under the local $U(1)$ with weight $3/2$ and $1/2$ respectively. The metric and the fourth rank AST are inert under both transformations. The scalars can be parameterized by a zweibein \cite{schwarz1}
\begin{equation}
V= \left(\begin{array}{cc} V_+^1&V_-^1\\
V_+^2&V_-^2\end{array}\right) = {1\over \sqrt{2i
\rho_2}}\left(\begin{array}{cc}
  \rho e^{i\phi}& \bar{\rho}e^{-i\phi}\\
 e^{i\phi}& e^{-i\phi}\end{array}\right).
\end{equation}

The group $SL(2,R)$ acts by matrix multiplication from the left and  
the local
$U(1)$ acts from the right and induces a shift $\phi\to \phi +\alpha$. We will
use  the   $U(1)$  gauge invariance to set $\phi=0$.  A compensating
gauge transformation accompanies an $SL(2,R)$ transformation in
order to
maintain the gauge condition.
In this case a  $SL(2,R)$ transformation of   the
complex scalar and  the compensating $U(1)$ transformation are  given by
\begin{equation}
\rho \to {a\rho+b\over c\rho+d},\quad e^{i\alpha}=\left({c\bar \rho+d\over c \rho+d}\right)^{\half}.
\end{equation}
It is useful to define a scalar field strength $P_\mu$ and a complex combination of the AST $H$ which have $U(1)$ weight 2 and 1 respectively.
\begin{equation}
\label{pdef}
P_\mu  =  -\epsilon_{ab} V_+^a \partial_\mu V_+^b=  {i\over 2}
{\partial_\mu
\rho\over
\rho_2},\quad H_{+\mu\nu\rho}=V_{+a}\partial_{[\mu}B^a_{\mu\rho]}. 
\end{equation}

\section{D-instanton in IIB supergravity}

The  classical D-instanton  is a solution of the euclidian  IIB  
supergravity
\cite{ggp}  which  preserves half the euclidian supersymmetry.    Only  
the metric
and the scalar fields are nontrivial in this background so the   
nontrivial part
of the action is
given by
\begin{equation}\label{action}
S=\int d^{10}x \sqrt{-g}\left\{ R-{1\over
2\rho_2^2}\partial_\mu\rho \partial^\mu \bar{\rho}\right\}.
\end{equation}
The BPS  constraint on the solution requires that
\begin{equation}\label{susyvar} \delta\lambda^*=i\gamma^\mu P_\mu\epsilon+..=0,\quad
\delta\psi_\mu=D_\mu\epsilon+..=0.
\end{equation}
The classical D-instanton solution is given,  in the Einstein frame,  by
setting
$g^E_{\mu\nu}=\eta_{\mu\nu}$.   If  the \RR\ scalar is expressed as  
the sum of
  a constant term and  an euclidian fluctuation, $\rho_1=  
\chi+if(x)$,  the BPS
condition (\ref{susyvar})  reduces to
$de^{-\phi}=-df$.
The  dilaton equation of motion is  solved by
\begin{equation}\label{solution}e^{\phi(x)-\phi_{\infty}}= 1+ {Ce^{-\phi_\infty}\over
|x-y|^8},
\end{equation}
where   $y$ is the position of the D-instanton in
space time.
 The
string coupling constant is identified with the expectation value of the
dilaton
at $|x|\to \infty$,  $g = e^{\phi_{\infty}}=1/\rho_2$.   The constant $C$ is
quantized
and given by $C= 2N/\pi^{3/2}$ where $N\in Z$. The  action of  a charge
$N$ D-instanton is  given by $S=2\pi i N \rho$.   The charge $N$  D-instanton
configuration preserves sixteen of the   thirty-two
supersymmetries. The broken supersymmetries generate fermionic zero
modes in  the
instanton background which  have
to be integrated over and they induce new
vertices which
soak up the zero modes
\begin{equation}\label{instvertex}
\int d^{10}y\;d^{16}\epsilon_0\; \langle
\psi^1(x_1)\cdots
\psi^n(x_n)\rangle_N = e^{2\pi i N \rho} \int d^{10}y\;d^{16}\epsilon
 \;\langle
\psi\rangle_{s_1}\cdots\langle \psi\rangle_{s_n},
\end{equation}
where $ \langle \psi\rangle_{s_i}$ defines a tadpole of a  field of IIB
supergravity which soaks up $s_i$ fermionic zero modes and $\sum_i
s_i =16$.

\section{Instanton induced interactions}\label{secinter}

In closed string perturbation theory processes in the presence of
 D-instantons are formulated  by including worldsheets with boundaries,
 where the boundary is mapped to a point in spacetime corresponding 
to the position of the instanton. Because the boundary conditions relate 
leftmoving and rightmoving worldsheet fields, a D-instanton  preserves 
 half of the closed string supersymmetry, i.e is a BPS state. 
The broken 
spacetime supersymmetry generators have the form of zero momentum open 
string fermion vertex operators and have a natural interpretation as
 generating fermionic zero modes in the instanton background.
 The simplest open-string world-sheet that arises in a D-brane
process is the
disk diagram.

\begin{figure}[h]
\begin{center}
\epsfig{figure=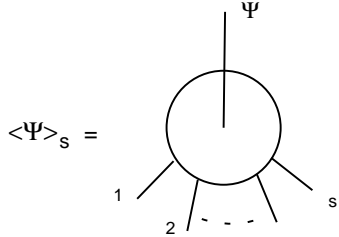,width=.4\textwidth}
\caption{An on-shell closed-string state, $\Psi$,   coupling to $s$
open-string fermions on a disk with Dirichlet boundary  conditions.}
\end{center}
\end{figure}

A tadpole diagram with one fermionic zero mode attached couples to the dilatino and the calculation in \ref{appa} gives
\begin{equation}
  \label{oneferm}
 \langle \lambda \rangle_1= \langle c F_{-{ 1\over 2}}(x)c\tilde{c}V_{(\lambda)} (z,\bar{z})\rangle= \bar \zeta_\lambda
\epsilon_0,
\end{equation}
where $\zeta_\lambda^a= \gamma^\mu_{ab} (\zeta_{\mu}^{b\; RN   } + i
\zeta_{\mu}^{b\;NR})$ is the holomorphic combination of the two
dilatinos.
Similarly one can evaluate the one point functions of the complex combination  of the \NSNS and \RR two-forms,   $B = B_{NS}+ i
C^{(2)}$ with two fermionic zero modes attached $\langle B\rangle_2$ and the tadpole $\langle \psi \rangle_3$ holomorphic gravitino with three fermionic zero modes attached. Of particular interest is the disk with four fermionic zero modes attached which couples to the graviton and the calculation in \ref{appb} gives 
\begin{eqnarray}
\langle h \rangle_4 =  \langle cF_{\half}(x_1)\int dx_2 dx_3 dx_4 
F(x_2)_{\half}
F_{-\half}(x_3) F_{-\half}(x_4)c\tilde{c}  V^{NN} (z,\bar{z}) \rangle\nonumber\\
=  \bar\epsilon_0\gamma^{\rho\mu\tau }\epsilon_0 \;
 \bar\epsilon_0\gamma^{\lambda\nu\tau }\epsilon_0 \; \zeta_{\mu\nu}k_\rho
 k_\lambda.\nonumber
\end{eqnarray}
The results of the calculation in the appendix of  tadpoles for the massless fields which absorb up to four fermionic zero modes are summarized  with their weight under $U(1)$ and the coupling  in the following table
\begin{table}[htb]
\begin{center}
\begin{tabular}{llll}
\hline
Field  & U(1)-wt  & no. z.m. & tadp\\
\hline
$\langle \lambda\rangle$  & 3/2  &  1& $\bar{\lambda}\epsilon$\\
$\langle H\rangle $& 1  & 2& $\bar{\epsilon}\gamma^{\mu\nu\rho}\epsilon H_{\mu\nu\rho}$\\
$\langle \psi\rangle$ & 1/2 & 3&$\bar{\epsilon}\gamma^{\mu\nu\rho}\epsilon \bar{\psi}_\rho\gamma_\mu\epsilon k_\nu$\\
$\langle h\rangle$ & 0 & 4&$\bar\epsilon\gamma^{\rho\mu\tau }\epsilon 
 \bar\epsilon\gamma^{\lambda\nu\tau }\epsilon \zeta_{\mu\nu}k_\rho$\\
\hline
\end{tabular}
\end{center}
\end{table}

There are a whole variety  of instanton induced vertices which are in fact related to each other by supersymmetry \index{effective action}.
The simplest case is given by a sixteen fermion term which is induced by sixteen disks with one holomorphic dilatino $\lambda$ and one zero mode $\epsilon$ attached to  each disk.
\begin{equation}
\label{lamcont}
  \int d^{10} y  \int
d^{16}\epsilon_0 \prod_{i=1}^{16} \langle \lambda\rangle_i =  \epsilon^{a_1\cdots a_{16}} \lambda^{a_1}\dots
\lambda^{a_{16}}.
\end{equation}
Among many other terms we focus on  a $R^4$ term which arises from four disks with a graviton and four fermionic zero modes attached to each disk. 
\begin{eqnarray}\label{fourgravss}
A_4  (\{\zeta_h^{(r)} \}) &=& \int d^{10} y  d^{16} \epsilon_0 \langle
h^{(1)}\rangle_4  \langle h^{(2)}\rangle_4  \langle h^{(3)}\rangle_4  \langle
h^{(4)}\rangle_4 \nonumber \\
&=&   \int d^{10} y  d^{16} \epsilon_0 \prod_{r=1}^4 \left(  \bar \epsilon_0
\gamma^{\mu_r\sigma_r\rho}\epsilon_0\ \bar \epsilon_0 \gamma^{\nu_r\tau_r
\rho}\epsilon_0 \zeta^{(\mu_r} \tilde \zeta^{\nu_r)} \
k_r^{\sigma_r}  k_r^{\tau_r} \right),\nonumber \\
 &=  &   \int d^{10}y \left(\hat{t}^{i_1j_1\cdots i_4j_4}\hat{t}_{m_1n_1\cdots
  m_4n_4}-{1\over 4}\epsilon^{i_1j_1\cdots j_4j_4}\epsilon_{m_1n_1\cdots
  m_4n_4}\right)\nonumber\\
&&\quad\quad \quad\quad\quad\times  R_{i_1j_1}^{m_1n_1}  R_{i_2j_2}^{m_2n_2}R_{i_3j_3}^{m_3n_3}\nonumber
\end{eqnarray}
 
The close connection between     $\Rfour$ and $\lam16$ terms is
obviously a
consequence of their  common
superspace \index{superspace} origin which can be seen already in the rigid limit of the
linearized theory.   Consider a  linear  superfield \index{superfield}
$\Phi(x,\theta)$  (where
$\theta$ is a complex chiral $SO(9,1)$ Grassmann spinor)  that
satisfies the
holomorphic constraint $ D^* \Phi=0$ and the on-shell condition
$D^4    \Phi
=  D^{*4} \Phi^*$ \cite{howewest} where
\begin{equation}
\label{covderiv}
 D_A = {\partial \over \partial \theta^A} +2i (\Gamma^\mu  \theta^*)_A
\partial_\mu, \qquad   D^*_A = - {\partial\over \partial\theta^{*A}}
\end{equation}
(recall that $\bar \theta = \theta\Gamma^0 $ does {\it not}
involve complex
conjugation) are the holomorphic and anti-holomorphic covariant
derivatives that
anticommute with
the rigid supersymmetries
\begin{equation}
\label{susys}
Q_A ={\partial \over \partial \theta^A}, 
\qquad   Q_A^* = - {\partial \over \partial
\theta^{*A}} +  2i
(\bar \theta \Gamma^\mu  )_A \partial_\mu .                                
\end{equation}
The field $\Phi$ has an expansion in powers of $\theta$ (but not $
\theta^*$), describing the 256 fields in an on-shell supermultiplet. 
\begin{eqnarray}
\label{phidef}
\Phi  &=& \rho_0 + \Delta\nonumber \\
 &= &\rho_0 + a  - {2i \over g}\thbar\lambda
  - {1 \over 24 g} \thbar\Gamma^{\mu\nu\sigma}\theta G_{\mu\nu\sigma}
+{i\over  6 g}\thbar\Gamma^{\mu\nu\sigma}\theta
\bar\theta\Gamma_{\nu}\partial_{\sigma}\psi_{\mu}  \nonumber \\
 &&\quad -{i \over 48
g}\thbar\Gamma^{\mu\nu\eta}\theta\thbar\Gamma_{\eta}^{\ \sigma\tau}\theta
R_{\mu\nu\sigma\tau}+
\cdots,
\end{eqnarray}
where $\Delta$ is the linearized fluctuation
around   a flat background with a constant scalar, $\rho_0 = \rho -  a =
C^{(0)}_0 +i g^{-1}$. Note that because of the constraint $D^4    \Phi
=  D^{*4} \Phi^*$ the expansion terminates at $\theta^8$.  The coefficients of the component fields  
are consistent
with the conventions used in \cite{schwarz1}.  The terms  indicated  
by $\cdots$
 fill in the remaining members of the ten-dimensional $N=2$ chiral
supermultiplet, comprising (in symbolic notation)   $\partial dC^{(4)}$,
$\partial
\psi_{\mu\nu}^*$,  $ \partial ^2 G^*_{\mu\nu\sigma}$,  $\partial  
^3\lambda^*$
and $ \partial^4\rho^*$.
 The linearized
supersymmetric on-shell action  has the form \cite{ggk}
\begin{equation}\label{actdef}
S' =   {\rm Re } \int d^{10}x  d^{16} \theta \,  g^4  F[\Phi],
\end{equation}
which is manifestly invariant under the rigid supersymmetry
transformations,
(\ref{susys}).
The various  component interactions contained in (\ref{actdef})   
are obtained
\ from the $\theta^{16}$ term in the expansion,
\begin{equation}
\label{expands}
F[\Phi] = F(\rho_0) + \Delta {\partial \over \partial \rho_0}  
F(\rho_o) + \half
 \Delta^2 \left( {\partial \over \partial \rho_0}\right)^2  F(\rho_o)   +
\cdots.
\end{equation}
All terms are F-terms, i.e they are realized as integrals over half the superspace.

\section{Modular Invariance}
The fact that the Lorentz index structure of the  $t_8t_8 R^4$ term in the instanton induced effective action is of the same form as the tree level \cite{wittengross} and one loop \cite{greenschwarziii}  perturbative terms  together with invariance of the graviton under the $SL(2,Z)$ duality of IIB \index{SL(2,Z)} string theory make it very plausible  to  conjecture for the exact nonperturbative $R^4$ term \cite{greengut}. 
\begin{equation}\label{r4mod}
S_{R^4}= \int d^{10}x\;  t_8t_8\ R^4\; f(\rho,\bar{\rho}),
\end{equation}
where the nonholomorphic modular function \index{modular invariance} $f$ is given by
\begin{equation}\label{modularguess}
  f(\rho,\bar{\rho})=  \sum_{(p,n )\neq
(0,0)}{\rho_2^{3/2}\over |p+n\rho|^3},
\end{equation}
In the string frame the large $\rho_2$ expansion gives
\begin{eqnarray}\label{expan}
\rho_2^{1/2} f (\rho, \bar \rho) &=&
2\zeta(3)(\rho_2)^{2} +
{2\pi^2\over
3}  + 4\pi^{3/2} \sum_{N}(N\rho^{ }_2)^{1/2}\sum_{N|\hat{m}}{1\over {\hat{m}}^2}\nonumber\\
&  
\times&  \left(e^{2\pi i N
\rho} + e^{-2\pi i N \bar \rho} \right) \left(1 + \sum_{k=1}^\infty
(4\pi N
\rho_2)^{-k} {\Gamma( k -1/2)\over \Gamma(- k -1/2) k!} \right)\nonumber \\
\end{eqnarray}
There are several interesting properties  of this expansion. There are  terms corresponding  to  tree level, one loop and instanton contributions. The tree and one loop terms exactly agree with results in string perturbation theory. The fact that no higher perturbative corrections appear in (\ref{expan}) together with the fact that all such terms have the form of integral over half the superspace, lead to the  conjecture that the $R^4$ and related terms   are protected from higher loop perturbative corrections by a nonrenormalization theorem. Recently substantial evidence for the validity of such a theorem has been obtained \cite{anton,berko}.
It is also interesting to note that it is possible to infer the value of certain integrals which are important for the calculation of the Witten index for bound states of D0-branes \cite{sethia} from the modular function $f$ \cite{ggd0}.
The nonholomorphic function $f$ is invariant under $SL(2,Z)$ modular transformations; this is  because the graviton and hence $t_8t_8 R^4$ is invariant under $SL(2,Z)$. For other terms which are related to $R^4$ like the $\lambda^{16}$ the modular function has to be generalized, since such terms transform with a nontrivial weight under the induced $U(1)$ R-symmetry. For  a  term in the effective action which transforms with weight $2k$ a natural generalization of (\ref{modularguess}) is \footnote{Similar considerations can be found in \cite{partouche}}, 
\begin{equation}
f^{(k)}=\sum_{(m,n)\neq (0,0)}{ \rho_2^{3/2}(m+n\rho)^{2k}\over |m+n\rho|^{2k+3}}.
\end{equation}
This function transforms under $SL(2,Z)$ with a $U(1)$ weight $k$
\begin{equation}
f^{(k)}({a\rho+b\over c\rho+d})= \left({c\bar{\rho}+d\over c\rho+d}\right)^k f^{(k)}(\rho).
\end{equation}
The functions $f^{(k)}$  with different $k$ are related to $f^{(0)}$ by 
\begin{equation}\label{moddef}
f^{(k)}= \rho_2^k D_{2(k-1)}\cdots D_2 D_0 f^{(0)},
\end{equation}
where 
\begin{equation}
D_l= i({\partial\over \partial \rho}- {l\over \rho-\bar{\rho}}),
\end{equation}
is the  nonholomorphic covariant derivatives which maps modular forms of weight $l$ into modular forms of weight $l+2$.

Comparing the  relations in (\ref{actdef}) with (\ref{moddef}) it becomes clear that the linearized superspace incorporated the action of the ordinary derivatives $\partial/\partial\rho$ by expanding the function $F[\Phi]$, but does not provide a covariantization. One indication that the full nonlinear superspace should produce a structure like (\ref{moddef})  can be obtained by considering the perturbative fluctuations around the instanton configurations which are supressed by powers of $\rho_2$. For the $\lambda^{16}$ term the lowest order contribution in the charge one instanton sector comes from sixteen disks with one dilatino and one fermionic zero mode attached. Because the genus of the disk is $-1$ and the insertion of a closed string vertex operator contributes one power of $g$ such a term contributes to order $g^0$. A  term where two dilatini are on the same disk with two fermionic zero modes attached is of order $g$ and corresponds to a perturbative fluctuation. There should be a contact term which contributes when the vertex operator of the two dilatini come close to each other.
Such contact terms are suppressed by $1/\rho_2=g$ and should be responsible for the covariantization of the leading D-instanton terms. Such terms with more than one closed string vertex operator on a disk together with disks with handles attached should produce the perturbative fluctuations around the instanton configuration \ref{expan}.
\section{One loop in eleven dimensions}
In this section we briefly review the relation of    $\Rfour$ term (\ref{r4mod}) by evaluating the four-graviton one loop amplitude  of
eleven-dimensional supergravity compactified  on a torus,  $T^2$, in the
directions $9$, $11$.     The volume of  the torus is    ${\cal V}=  
R_9 R_{11}$
and its complex
structure,  $\Omega
= \Omega_1 + i\Omega_2$, may be expressed in terms of the components
eleven-dimensional metric   $G_{\mh \nh}$ ( $\mh = 0, \cdots, 9,11$)  as
$\Omega_2 = G_{9\; 9} / G_{11\; 11} =  R_9/R_{11}$ and  $\Omega_1 = G_{9\;
11}/G_{11\;11} = C^{(1)}$, where $C^{(1)}$ is the component of the
IIA one-form
along the direction $x^9$. $M$-theory on $T^2$ is identified with $IIB$ on $S^1$ via setting $\Omega=\rho$.
 The expression for the amplitude,   obtained rather efficiently  by
considering a first-quantized super-particle in the light-cone gauge
\cite{ggv},  has the structure,
\begin{equation}
\label{loopgen}
A_{R^4}   =
\int d^9 p  {1
\over {\cal V}} \sum_{m,n}\int \prod_{r=1}^4 d\tau_r  \tr(V^{(1)}_h(k^1)
V_h^{(2)}(k^2)
V_h^{(3)} (k^3)V_h^{(4)} (k^4) ),
\end{equation}
where $V^{(r)}_h(k^r)$ is the graviton vertex operator for the $r$th
graviton with momentum $k^r$ and is
 evaluated at a proper time $\tau_r$ around the loop (and dimensional quantities 
are measured in units
of the eleven-dimensional Planck scale).   The  trace is
over the
fermionic zero modes and the loop momentum in the directions
$0,\cdots,8$
and  the  integers
$m$ and $n$
 label the Kaluza-Klein momenta in the directions $9$, $11$ of $T^2$.
  The trace over the fermionic modes gives an overall factor, $K$, which
contains eight powers of the external momenta and is the linearized
form of
$\Rfour$.    The $\Rfour$ term in (\ref{loopgen}) is  obtained by
setting the
momenta equal to zero in the remainder of the loop integral.  It is
convenient
to perform a double Poisson resummation in order to rewrite the sum
over the
Kaluza--Klein momenta  as a sum over windings, $\mm$ and $\nn$, of
the  loop
around the cycles of $T^2$, giving,
\begin{eqnarray}
\label{loopgrav}
{\cal V } A_{R^4}  &=&   K {\cal V} \sum_{\mm,\nn}\int {d \tau} \tau^{\half}
e^{- \tau {\cal V}{ |\mm+\nn\Omega|^2\over \Omega_2}}  \nonumber\\
&=&  K{\cal V} C+\pi^2  K{\cal V}^{-\half} f(\Omega,\bar \Omega),
\end{eqnarray}
where
\begin{equation}
\label{modfun}
f(\Omega,\bar{\Omega})=  \pi^{-2} \Gamma(3/2)
\sum_{(m,n)\neq(0,0)}{\Omega_2^{3/2}\over
|m+n\Omega|^3}
\end{equation}
is a modular function.
The cubic ultraviolet divergence of $A_{R^4}$ is contained in the zero
winding term,   $\mm = 0 = \nn$,  which  has the divergent
coefficient, $C$.
It was argued in \cite{greenhove} that  in any regularization that is
consistent with
T-duality between the IIA and IIB string theories in nine
dimensions  this has
to be replaced by a regularized finite value, $C= \pi/3$ \cite{ggv}.
 Presumably a
microscopic description of M-theory (such as   matrix theory)
would reproduce
this value.    In the limit of zero volume, ${\cal V} \to 0$, this term
disappears
and  only the term with coefficient ${\cal V}^{-1/2}$ survives. Similar results for the $\lambda^{16}$ and related terms and the corresponding modular functions $f_k$  have been obtained in \cite{ggk} by evaluating different one  loop amplitudes of $M$ compactified  on $T^2$.

\acknowledgements 

I would like to thank M.B. Green, H.H. Kwon and P. Vanhove for  collaboration  and  discussions on matters presented in this talk. I gratefully acknowledge the  partiall support by a DOE grant
 DE-FG02-91ER40671, NSF grant PHY-9157482 and a James S.
 McDonnell grant 91-48.

\appendix 
\section{One point functions in the upper half plane}
We give two examples of the calculation of one point functions on the half plane which give the tadpoles $\langle \Psi \rangle$ of closed string fields used in section \ref{secinter}.

\subsection{Dilatino}\label{appa}

Attaching one fermionic zero mode to the boundary leads to a
non-vanishing one-point function for a combination of the two dilatinos
from the \NSR\
\begin{eqnarray}
 \langle\lambda^{NR}\rangle_1&=& \langle
c\bar{c}
V_{(-1,-\half)}(v,z)cV_{-\half}(\epsilon,x_1)\rangle\nonumber\\
&=& v_{\mu a}\epsilon_b\langle c\bar{c}e^{-\phi}\psi^\mu e^{-\half\bar{\phi}}\bar{S}^a
  e^{ikX}(z,\bar{z})ce^{-\half \phi}S^b(x_1)\rangle\nonumber\\
&=& i v^{\mu}_a\epsilon_b
(z-x_1)(\bar{z}-x_1)(z-\bar{z}){\gamma_\mu^{ab}\over (z-x_1)(\bar{z}-x_1)(z-\bar{z})}\nonumber\\
&=& i\bar{\epsilon}\lambda\label{dilatino1}
\end{eqnarray}
and similarly for the  \RNS\ dilatino
\begin{eqnarray}
\langle\lambda^{RN}\rangle_1&=&  \langle
c\bar{c}
V_{(-\half,-1)}(v,z)cV_{-\half}(\epsilon,x_1)\rangle= \bar{\epsilon}\lambda.\label{dilatino2}
\end{eqnarray}
The calculation of such correlation functions on the upper half plane is most easily done by doubling. One uses the boundary conditions for the D-instanton  to map  the antiholomorphic fields  of the vertex operators living in the upper half plane to holomorphic fields in the lower half plane and evaluates the correlation function using holomorphic correlators in the complex plane.
Putting (\ref{dilatino1}) and (\ref{dilatino2}) together the coupling
of the dilatinos is given by
\begin{equation}
  \langle \lambda \rangle_1 =   (\zeta_{a\mu}^{RN   } + i   \zeta_{a\mu}^{NR})
(\gamma^{\mu})^a_b  \epsilon^b_0. 
\end{equation}

\subsection{\NSNS\ antisymmetric tensor}\label{B:NSNS}

Attaching two fermionic zero modes produces a coupling of the
D-instanton to the antisymmetric rank 2 tensor fields from the \NSNS\
and \RR\ sector. For the \NSNS\ antisymmetric tensor the relevant amplitude is
\begin{eqnarray}
 \langle B^{NN}_{\mu\nu}\rangle_2&=&  \langle
c\bar{c}
V_{(0,-1)}(v,z)cV_{-\half}(\epsilon,x_1)\int dx_2
V_{-\half}(\epsilon,x_2)\rangle\nonumber \\
&=&\zeta^{NN}_{\mu\nu}\epsilon^1_a\epsilon^2_b \left\langle c\bar{c} e^{-\phi}\psi^\mu(\bar{\partial}
  X^\nu+ik_\lambda\bar{\psi}^\lambda\bar{\psi}^\nu)e^{ikX}(z,\bar{z})
c e^{-\half \phi}S^a(x_1)\right.\nonumber\\
&&\quad\quad\quad\times \left.\int dx_2e^{-\half \phi}S^b(x_2)\right\rangle
\end{eqnarray}
Note that the $\bar{\partial} X^\mu$ cannot be contracted into
anything, the $\bar{\psi}^\lambda\bar{\psi}^\nu(\bar{z})$ in
$V_{(0,0)}$ acts as a current  $\bar{j}^{\lambda\nu}$ insertion which
can be evaluated,
\begin{eqnarray}
\langle B_{\mu\nu}\rangle_2&=& ik_\lambda\zeta^{NN}_{\mu\nu}\epsilon_1^a\epsilon_2^b\int
  dx_2 {(z-\bar{z})(\bar{z}-x_1)\over
  (z-x_2)(x_1-x_2)}\nonumber\\
&&\quad \times \left({\gamma^{\lambda\nu\mu}_{ab}\over
  \bar{z}-x_1}-{\gamma^{\lambda\nu\mu}_{ab}\over
  \bar{z}-x_2}+{\eta^{\lambda\mu}\gamma^\nu_{ab}-\eta^{\nu\mu}\gamma^\lambda_{ab}\over z-\bar{z}}\right)\nonumber\\
  &=&
  ik_\rho\zeta^{NN}_{\mu\nu}\bar{\epsilon}\gamma^{\mu\nu\rho}\epsilon\int dx_2
  {(z-\bar{z})\over (z-x_2)(\bar{z}-x_2)}\nonumber\\
&=&\pi H^{NN}_{\mu\nu\rho}\bar{\epsilon}\gamma^{\mu\nu\rho}\epsilon.\label{Bmunu1pt}
\end{eqnarray}
Where we defined $ H^{NN}_{\mu\nu\rho}=ik_\rho\zeta^{NN}_{\mu\nu}$.
Only the first two terms proportional to $\gamma^{\mu\nu\rho}$
contribute because the term proportional to just one $\gamma$ matrix
in the first line of (\ref{Bmunu1pt})  $\bar{\epsilon}_1\gamma^\mu \epsilon_2$
vanishes after the spinor wavefunctions $\epsilon^1$ and $\epsilon^2$ are antisymmetrized. In the last line we used the fact that
\begin{equation}\label{integral1}
  \int dx {(z-\bar{z})\over(z-x)(\bar{z}-x)}=\pi
\end{equation}
which is true for any $z$ with non vanishing imaginary part by the residue-theorem.

\subsection{\RR\ antisymmetric tensor}

For the \RR\ antisymmetric tensor the relevant amplitude is given by
\begin{eqnarray}
   \langle B^{RR} \rangle_2&=&\langle
c\bar{c}
V_{(-\half,-\half)}(F^{(1)},z)cV_{-\half}(\epsilon,x_1)\int dx_2
V_{-\half}(\epsilon,x_2)\rangle\nonumber \\
&=&{i\over 4}k_{\mu_1}\zeta^{RR}_{\mu_2\mu_3}\epsilon^1_c\epsilon^2_d\left\langle c\bar{c}
  e^{-\half\phi}e^{-\half\bar{\phi}}S^a\gamma^{\mu_1\mu_2\mu_3}_{ab}\bar{S}^b e^{ikX}\right.\nonumber \\
&&\left. \quad \times c e^{-\half \phi}S^c(x_1)\int dx_2e^{-\half
    \phi}S^d(x_2)\right\rangle\nonumber\\
&=&-{1\over 4} k_{\mu_1}\zeta^{RR}_{\mu_2\mu_3}\epsilon^1_c\epsilon^2_d\int dx_3\nonumber\\
&& \times \left((\bar{z}_1-x_2)
 {\tr( \gamma^\rho \gamma^{\mu_1\mu_2\mu_3})\gamma_{\rho}^{ cd}\over
 (\bar{z}-x_2)(x_1-x_2)}-(z-\bar{z}){(\gamma^\rho\gamma^{\mu_1\mu_2\mu_3}\gamma_\rho)^{cd}\over (z-x_2)(\bar{z}-x_2)}\right)\nonumber\\
&=& i\pi H^{RR}_{\mu\nu\rho}\bar{\epsilon}\gamma^{\mu\nu\rho}\epsilon\label{B-RR}
\end{eqnarray}
Again $H^{RR}_{\mu\nu\rho}=ik_\rho \zeta^{RR}_{\mu\nu}$.
Note that the first term vanishes because of the trace of
$\gamma$-matrices and Bose-Fermi symmetry of $\epsilon_1,\epsilon_2$. To evaluate
the second term we used the
identity
$\gamma^\rho\gamma^{\mu_1\mu_2\mu_3}\gamma_\rho=-4\gamma^{\mu_1\mu_2\mu_3}$
for ten dimensional $\gamma$-matrices and 
the  integral  (\ref{integral1}). 

Putting together (\ref{Bmunu1pt}) and (\ref{B-RR}) we see that the
following  couples to the D-instanton with two fermionic zero
modes attached. 
\begin{equation}
  \langle B \rangle_2 =\bar\epsilon_0\gamma^{\mu\nu\rho} \epsilon_0 ik_{[\mu}
(\zeta^{NN}_{\nu\rho]} + i \zeta^{RR}_{\nu\rho]})
\end{equation}

\subsection{Graviton}\label{appb}

Attaching four fermionic zero modes gives a coupling of the self-dual
\RR\ four form and the graviton to the D-instanton. We focus here on the
graviton where the relevant amplitude is given by
\begin{eqnarray}
\langle h\rangle_4&=&   \langle
c\bar{c}
V_{(0,0)}(\zeta,z)cV_{-\half}(x_1)\int dx_2 dx_3 dx_4
V_{-\half}(x_2)
V_{-\half}(x_3)
V_{-\half}(x_4)\rangle\nonumber\\
&=&\zeta_{\mu\nu}\epsilon_1^a\epsilon_2^b\epsilon_3^c\epsilon_4^d\left\langle
c\bar{c}(\partial
  X^\mu+ik_\rho\psi^\rho\psi^\mu)(\bar{\partial}
  X^\nu+ik_\lambda\bar{\psi}^\lambda\bar{\psi}^\nu)e^{ikX}(z,\bar{z})\right.\nonumber\\
&&\times  \left. ce^{- {\phi\over 2}}S^a(x_1)\int dx_2 dx_3 dx_4 e^{- {\phi\over 2}}S^b(x_2)e^{- {\phi\over 2}}S^c(x_3) e^{- {\phi\over 2}}S^d(x_4)\right\rangle.\nonumber
\end{eqnarray}
Using doubling and the well known correlation function for the superghosts and spin fields, in particular 
\begin{eqnarray}
  &&\left\langle
  j^{\rho\mu}(z)j^{\lambda\nu}(\bar{z})S^a(x_1)S^b(x_2)S^c(x_3)S^d(x_4) 
\right\rangle\nonumber\\
&=&\sum_{i=1}^4{M^{\rho\mu}(i)\over
  z-x_i}\sum_{j=1}^4{M^{\lambda\nu}(i)\over
  \bar{z}-x_i}{x_{14}x_{23}\gamma^\mu_{ab}\gamma_{\mu\;
  cd}-x_{12}x_{34}\gamma^\mu_{ad}\gamma_{\mu\; bc}\over
  (x_{12}x_{13}x_{14}x_{23}x_{24}x_{34})^{3/4}}.
\label{MM}
\end{eqnarray}
The operator $M(i)$ acts on the spinor index associated with the i-th
spin field as a Lorentz transformation. As above only terms like
$\epsilon_i\gamma^{\mu_1\mu_2\mu_3}\epsilon_j$ are nonzero after antisymmetrization
over the spinor wavefunctions $\epsilon_i$.

For example the  contributions of the form
$\bar{\epsilon}_1\gamma^{\rho\mu\tau}\epsilon_2
\;\bar{\epsilon}_3\gamma^{\lambda\nu\tau}\epsilon_4$ is given
by  $M^{\rho\mu}$ acting on $\gamma^\tau_{ab}$ and
$M^{\lambda\nu}$ acting on $\gamma^\tau_{cd}$, which results in 
\begin{eqnarray}
  &&\bar{\epsilon}_1\gamma^{\rho\mu\tau}\epsilon_2
  \;\bar{\epsilon}_3\gamma^{\lambda\nu\tau}\epsilon_4\int
  dx_2dx_3dx_4{(z-\bar{z})(z-x_1)(\bar{z}-x_1)\over
  (x_{12}x_{13}x_{24}x_{34})}\nonumber\\
&&\quad\quad \quad\quad\quad \times \left( {1\over z-x_1}-{1\over
  z-x_2}\right)\left( {1\over \bar{z}-x_3}-{1\over
  \bar{z}-x_4}\right)\nonumber \\
&=& \bar{\epsilon}_1\gamma^{\rho\mu\tau}\epsilon_2
  \;\bar{\epsilon}_3\gamma^{\lambda\nu\tau}\epsilon_4\int
  dx_2dx_3dx_4 {(z-\bar{z})(\bar{z}-x_1)\over
  x_{12}x_{34}(z-x_2)(\bar{z}-x_3)(\bar{z}-x_4)}\nonumber\\
&=&\bar{\epsilon}_1\gamma^{\rho\mu\tau}\epsilon_2
  \;\bar{\epsilon}_3\gamma^{\lambda\nu\tau}\epsilon_4\int dx_2 {(z-\bar{z})\over
  (z-x_2)(\bar{z}-x_2)}\nonumber\\
&=&\pi \bar{\epsilon}_1\gamma^{\rho\mu\tau}\epsilon_2
  \;\bar{\epsilon}_3\gamma^{\lambda\nu\tau}\epsilon_4.
\end{eqnarray}
There are three  other contributions in (\ref{MM}) given by 
$\bar{\epsilon}_1\gamma^{\lambda\nu\tau}\epsilon_2
\;\bar{\epsilon}_3\gamma^{\rho\mu\tau}\epsilon_4$, $\bar{\epsilon}_1\gamma^{\lambda\nu\tau}\epsilon_4
\;\bar{\epsilon}_2\gamma^{\rho\mu\tau}\epsilon_3$ and  $\bar{\epsilon}_1\gamma^{\rho\mu\tau}\epsilon_4
\;\bar{\epsilon}_2\gamma^{\lambda\nu\tau}\epsilon_3$. 
After antisymmetrization over all $\epsilon_i,i=1,2,3,4$ these terms add up
to give 
like  the result for the graviton four point function
\begin{equation}
  \langle h\rangle_4=\pi  \zeta_{\mu\nu}k_\rho k_\lambda \bar{\epsilon}\gamma^{\rho\mu\tau}\epsilon \;\bar{\epsilon}\gamma^{\lambda\nu\tau}\epsilon
\end{equation}

\end{document}